\documentclass[%
preprint,
 amsmath,amssymb,
 aps,
 prfluids,
 showkeys
]{revtex4-2}

\usepackage{graphicx}
\usepackage{dcolumn}
\usepackage{bm}
\usepackage{float}
\usepackage{placeins}
\usepackage{xcolor}
\usepackage{soul}
\usepackage[percent]{overpic}
\usepackage{tikz}
\usepackage{multirow}
\usepackage{array}

\usepackage{soul}

\begin{document}

\preprint{APS/123-QED}

\title{Information-theoretic characterization of turbulence intermittency}

\author{Shreyashri Sarkar}
\affiliation{%
 Department of Aerospace Engineering, Indian Institute of Science, Bangalore, KA, India
}%
\author{Rishita Das}%
 \email{rishitadas@iisc.ac.in}
\affiliation{%
 Department of Aerospace Engineering, Indian Institute of Science, Bangalore, KA, India
}%

\date{\today}

\begin{abstract}
Small-scale intermittency is studied as the deviation of the probability distributions of pseudodissipation, dissipation and enstrophy in turbulence from those of a Gaussian random velocity field. This deviation is quantified using Kullback-Leibler (KL) divergence between the two distributions, directly measuring turbulence-induced intermittency separated from purely kinematic effects. 
Using direct numerical simulation data of forced isotropic turbulence over a wide range of Taylor Reynolds numbers ($Re_{\lambda}$), we characterize the $Re_{\lambda}$ dependence of small-scale intermittency via KL divergence and uncertainty via Shannon entropy, identifying distinct behavioral regimes.
Small-scale uncertainty exhibits a non-monotonic dependence on $Re_{\lambda}$: despite continuously growing variability, entropy decays above a certain Reynolds number, suggesting a fundamental change in the statistical nature of the small scales. 
Turbulence-induced intermittency grows logarithmically with Reynolds number in contrast to the commonly reported power-law scaling, implying that turbulence shows a diminishing growth rate of intermittency at higher Reynolds numbers. 
Finally, we uncover an emergent symmetry: turbulence dynamics is shown to generate nearly equal intermittency in dissipation rate and enstrophy, challenging the prevailing assumption of asymmetry between strain-rate and vorticity dynamics.
\end{abstract}

\maketitle

\section{Introduction}

In a turbulent flow, velocity increments at large scales are approximately Gaussian; however, as the scale decreases, their probability distributions progressively deviate from Gaussianity, exhibiting heavier tails and an increased likelihood of extreme events \cite{monin1975statistical,castaing1990velocity,gotoh2002velocity}.
In the limit of vanishingly small scale, velocity increments correspond to velocity gradients.
The irregular and sporadic spatio-temporal occurrence of extreme velocity gradients in a turbulent flow manifests as a specific non-Gaussian behavior that is referred to as small-scale intermittency \cite{KRS, lawson2015velocity, wilczek2014pressure, tsinober}.
This small-scale intermittency is thus defined as the deviation of dissipation rate, enstrophy, and pseudodissipation rate in a turbulent flow field from a non-intermittent Gaussian random field \cite{cao1999statistics, rosales2008anomalous, Gotoh22}.
Statistical characterization of the intermittent nature of these quantities is essential not only in advancing our fundamental understanding of turbulence, but also in turbulence modeling \cite{pope2001turbulent, durbin2018some}.
Understanding intermittency is a crucial step toward developing a comprehensive statistical theory of turbulent flows.
Traditionally, small-scale intermittency has been quantified using higher-order moments of these quantities \cite{ schumacher2014small, sreenivasan2021dynamics, yadon17, yakt18, schumacher2007asymptotic}.
Yakhot and Donzis \cite{yadon17, yakt18} and Schumacher et al. \cite{schumacher2007asymptotic} showed that intermittency measured this way begins to grow only beyond a transitional Taylor Reynolds number, $Re_{\lambda} \sim 10$, where $Re_{\lambda} \equiv u_{rms}\lambda/\nu$, with $\lambda$ as the Taylor micro-scale, $u_{rms}$ the root mean square of velocity fluctuations, and $\nu$ the kinematic viscosity. The moments scale as power laws of $Re_{\lambda}$, indicating a persistent, unbounded growth of intermittency with Reynolds number \cite{yakhot2017multiscaling, yadon17, schumacher2007asymptotic, elsinga}, nearly universally across diverse turbulent flows \cite{schumacher2014small}. Such analyses have further led to the inference that enstrophy is more intermittent than dissipation, a feature commonly considered to reflect an asymmetry in vorticity-strainrate dynamics of turbulent flows \cite{don18, bermejo2009geometry, zeff2003measuring, chen1997inertial, siggia1981numerical, kerr1985higher, buaria2022vorticity}.

While previous studies have predominantly relied on higher order moments of dissipation, enstrophy, and pseudodissipation to quantify intermittency, this approach has limitations. 
Moments of different orders show different scaling exponents
\cite{yakt18, yadon17, schumacher2007asymptotic, Elsinga_Ishihara_Hunt_2023, extremevents} and 
each order moment represents a specific geometric feature of the probability density function (PDF), and not the complete PDF.
Consequently, using a single order moment to quantify intermittency introduces ambiguity in its interpretation \cite{gbel}.
Moreover, higher-order moments are highly sensitive to rare events and require a large number of samples for statistical convergence.
Recently, Granero-Belinchón et al. \cite{gbel} have demonstrated that Kullback-Leibler (KL) divergence \cite{KLD}, an information-theoretic quantity that accounts for the deviation of the entire PDF instead of select moments, 
offers a precise measure of inertial-range intermittency.
KL divergence is also more robust to sampling errors \cite{bovolo2008_resolution_main, resolution_pix, resolution_fig} and has previously been employed to study transition in turbulent pipe flows \cite{raj2025detecting} and self-similarity and non-Gaussianity of velocity in turbulent boundary layers \cite{zhou2015properties, tsuji1999probability}.

In this work, we introduce KL divergence of dissipation, enstrophy or pseudodissipation as a unified, comprehensive measure of small-scale intermittency.
This measure captures the extent of heavy-tailed, non-Gaussian nature of turbulence small-scale quantities. For the purpose of this paper, intermittency refers to deviation from Gaussianity.
This intermittency is separated into two counterparts - kinematic and turbulence-induced. Kinematic intermittency is the intermittency arising from the kinematic definition of the quantity, present even in an uncorrelated Gaussian random field (GRF).
In contrast, turbulence intermittency is the deviation of the turbulence quantity from the Gaussian random field, representing the intermittency arising only from the turbulence dynamics.

We apply this framework of quantifying intermittency to direct numerical simulation (DNS) datasets of homogeneous isotropic turbulent flows of a wide-range of Taylor Reynolds numbers. 
We establish the scaling of turbulence-induced small-scale intermittency and the associated uncertainty (quantified via Shannon entropy) with increasing Reynolds number, 
clearly identifying the transition Reynolds numbers where the nature of small-scale turbulence changes markedly.
In addition, we uncover an emergent symmetry in the intermittent dynamics of strain-rate and vorticity arising from turbulence.
Overall, using this new framework we 
examine the intermittency and uncertainty of the small scales of turbulent flows, challenging some of the widely accepted views, revealing new insight and refining our understanding of small-scale turbulence.

The remainder of the paper is organized as follows. Section \ref{sec:info} introduces the information theoretic measures of interest in this study, and section \ref{sec:interm}  discusses the non-intermittent GRF, followed by definition of KL divergence-based measures of kinetic and turbulence intermittency, illustrated through an analytical example representative of small-scale turbulent quantities. 
The DNS datasets used in this work are described in section~\ref{sec:DNS}. Section~\ref{sec:pseudo} 
 presents the Reynolds numbers variation of the probability distributions of the small-scale quantities.
Finally, the entropy and intermittency of turbulence small scales and their Reynolds number dependence is characterized in sections~\ref{sec:entropy_results} and \ref{sec:interm_results}, respectively, concluding with a summary of the primary findings of the study.

\section{Information-theoretic measures}\label{sec:info}

Information theory is fundamentally based on the concept of Shannon entropy \cite{shannon,khinchin2013mathematical,lesne2014shannon}, which measures the uncertainty or information content of a random variable $X$ as
\begin{eqnarray}
    \mathrm{H}(X)=- \sum_{x \in \Pi_x}p(x)\log{(p(x))} = &-& \int_{\Pi_x} f(x)\log{\left(f(x)\right)}\; dx - \log{(\Delta x)}
    \label{eqn3}
\end{eqnarray}
where $p(x)$ is the probability and $f(x)$ is the PDF of a realization $x$ (out of all possible realizations $\Pi_{x}$) of the random variable $X$. 
Here, $p(x) \approx f(x)\Delta x$, for a finite, uniform spacing $\Delta x$ between consecutive realizations.
The relation between the discrete Shannon entropy (summation) and continuous differential entropy (integral) in Eq.~(\ref{eqn3}) is valid under the assumptions of small $\Delta x$ and Riemann integrability of the integrand \cite{cover}. By definition, Shannon entropy is always non-negative, since probability satisfies $ 0 \leq p(x) \leq 1$.
Throughout this work, all logarithms are natural $(\log_e \equiv \ln)$, and hence, all entropy-based measures are in units of nats.

Kullback-Leibler (KL) divergence or relative entropy of the PDF $f(x)$ of an intermittent random variable $X$, from a reference or baseline PDF $f_0(x)$, is given by \cite{KLD,gbel}
\begin{eqnarray}
    D_{f||f_0}(X) = \int_{\Pi_x} f(x)\log{\left(\frac{f(x)}{f_0(x)}\right)}dx \; = \sum_{x \in \Pi_x} p(x)\log{\left(\frac{p(x)}{p_0(x)}\right)},
    \label{eqn1}
\end{eqnarray} 
which is the amount of information lost when $f_{0}(x)$ is used to represent $f(x)$.
KL divergence is a non-negative asymmetric measure of the ``distance" between two PDFs, $f(x)$ and $f_0(x)$, that vanishes only when the PDFs are identical, i.e., $f(x)=f_{0}(x)$.
It quantifies the difference in the shapes of the two probability distributions, therefore, accounting for the contributions of all the moments of $X$ in each distribution. 
As before, the equality between the continuous and discrete forms of KL divergence holds for a small enough $\Delta x$ and a Riemann-integrable integrand, both of which are satisfied in our analysis.

\section{Measure of intermittency}\label{sec:interm}

To characterize the small-scale behavior of turbulence, we analyze velocity gradient $\left(A_{ij} \equiv \partial u_i/\partial x_j\right)$ quantities such as pseudodissipation rate $(\phi = A_{ij}A_{ij})$, dissipation rate $(\epsilon = S_{ij}S_{ij})$, and enstrophy $(\Omega = W_{ij}W_{ij})$ in a turbulent field. Here, $S_{ij} \equiv (A_{ij}+A_{ji})/2$ is the symmetric strain rate tensor and $W_{ij} \equiv (A_{ij}-A_{ji})/2$ is the skew-symmetric rotation rate tensor.
We investigate the uncertainty (using Shannon entropy) and intermittency (using KL divergence) of these quantities.
Following prior studies \cite{yadon17, yakt18, buaria2022vorticity, buaria2022scaling, Yeung_Donzis_Sreenivasan_2012, extremevents, kinematic, Elsinga_Ishihara_Hunt_2023}, mean-normalized fields, $X/\langle X \rangle$, where $X = \phi$, $\epsilon$ or $\Omega$, are used to investigate small-scale intermittency. Hereafter, mean-normalized quantities $X/\langle X \rangle$ are denoted as $X$ for notational simplicity and all information-theoretic measures are applied to these normalized variables.

\subsection{Gaussian Random Field}

A zero-mean divergence-free spatially-uncorrelated isotropic Gaussian random velocity field or GRF 
can be considered to be the non-intermittent baseline field \cite{cao1999statistics, Gotoh22}. 
Such an uncorrelated GRF represents the purely random Gaussian state, free of any dynamics; deviation of the turbulent state from this Gaussian state thus signifies the intermittency arising from turbulence dynamics.

\begin{figure}
\includegraphics[width=0.6\columnwidth]{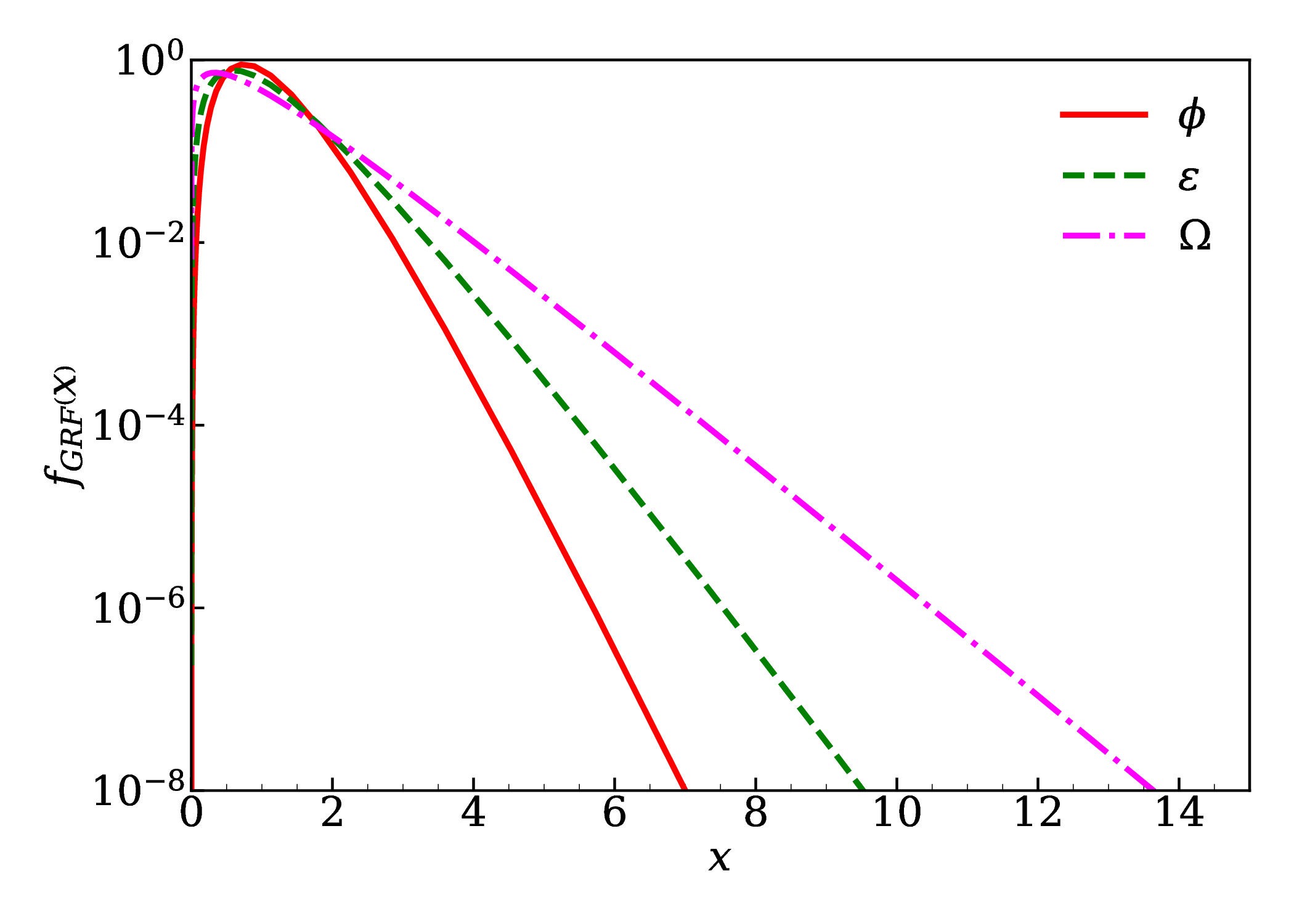}
\caption{\label{fig:1} PDFs of pseudodissipation rate $\phi$, dissipation rate $\epsilon$, and enstrophy $\Omega$, in non-intermittent baseline (Gaussian random) field, $f_{GRF}(X=x)$ for $X=\phi,\epsilon$ and $\Omega$.}
\end{figure}

In such a Gaussian field, not only the velocity ($u_i$) but also the velocity gradients (${A}_{ij}$) are Gaussian \cite{solak}. However, the quantities of our interest -- mean-normalized pseudodissipation, dissipation, and enstrophy, which are quadratic in ${A}_{ij}$, 
are not Gaussian. These quantities exhibit the following Gamma distributions as shown previously by Gotoh and Yang \cite{Gotoh22} (derivation in Appendix~\ref{appA}):
\begin{eqnarray}
f_{GRF}(x) =\frac{\left(\frac{n}{2}\right)^{\frac{n}{2}}}{\Gamma \left(\frac{n}{2}\right)}x^{\frac{n}{2}-1}\exp\left(-\frac{nx}{2}\right),
\label{eqn4}
\end{eqnarray}
where $x$ is a probable realization of $X = \phi, \epsilon$ or $\Omega$, and 
$\Gamma(\cdot)$ denotes gamma function. Here, $n$ denotes the number of independent terms in the definition of $X$, which is equal to $8, 5$ and $3$ for $\phi$,  $\epsilon$ and $\Omega$, respectively, in a three-dimensional incompressible velocity field \cite{Gotoh22, meneveau2011lagrangian}. 
All three PDFs are illustrated in Fig.~\ref{fig:1}.
Notably, the PDFs $f_{GRF}$ are independent of the standard deviation of the velocity field since $ \phi, \epsilon$ and $\Omega$ are mean-normalized (Appendix~\ref{appA}).
This suggests that $f_{GRF}$ serves as a universal Gaussian baseline with respect to which intermittency of any divergence-free isotropic field may be measured. 

\subsection{Kinematic Intermittency}

Interestingly, quadratic quantities such as pseudodissipation, dissipation, and enstrophy follow Gamma distributions (Eq.~\ref{eqn4}) even in a non-intermittent Gaussian velocity field, rendering them inherently non-Gaussian. 
This intrinsic deviation of a quantity from a Gaussian distribution is referred to as its kinematic intermittency.
The concept of kinematic intermittency -- arising from the nonlinear combinations of Gaussian velocity gradients -- was introduced by Shtilman et al. \cite{shtilman1993some} and Tsinober et al. \cite{tsinober, tsinober2019essence}. 
Here, kinematic intermittency of a variable $X$ is measured by the KL divergence of its PDF in a GRF with respect to that of a Gaussian PDF $f_G(x)$ of the same mean and variance (Eqs.~\ref{eqn1},\ref{eqn4}):
\begin{eqnarray}
D_{f_{GRF}||f_{G}}(X)&=\log{\left(\frac{\sqrt{n\pi}}{\Gamma\left(\frac{n}{2}\right)e^{\frac{n-1}{2}}}\right)} + \left(\frac{n}{2}-1\right)\psi\left(\frac{n}{2}\right)
\;\;\; \text{where} \; \psi(z) = (d\Gamma(z)/dz)/\Gamma(z)
\label{eqn4.5}
\end{eqnarray}
The term ``kinematic" stems from the fact that this intermittency originates purely from the definitions of these quantities -- governed solely by the number of independent nonlinear terms $n$ (Appendix~\ref{appA}), rather than any system dynamics.
The kinematic intermittency of pseudodissipation, dissipation, and enstrophy are mutually distinct -- increasing with decreasing $n$, 
such that $D_{f_{GRF}||f_{G}}(\phi) < D_{f_{GRF}||f_{G}}(\epsilon) < D_{f_{GRF}||f_{G}}(\Omega)$ ($0.089 < 0.147 < 0.261$). This ordering is also evident in Fig.~\ref{fig:1}, where $\Omega$ exhibits the heaviest tailed Gamma PDF, followed by $\epsilon$ and $\phi$. This shows that enstrophy is the most intermittent even in a GRF and this intermittency is kinematic.
This intermittency persists in these quantities independent of the underlying dynamics of the system.

\subsection{Turbulence intermittency}

In a turbulent flow field, dissipation, enstrophy and pseudodissipation deviate further from the Gamma distribution of the GRF, exhibiting increasingly heavy-tailed probability distributions at higher Reynolds numbers \cite{Gotoh22}.
This additional intermittency, present in turbulence but absent in a Gaussian velocity field, arises from the dynamics of turbulence and is therefore referred to as turbulence-induced intermittency.
It can be quantified by measuring the deviation of a turbulent quantity from that in a GRF, using KL divergence:
\begin{eqnarray}
    D_{f||f_{GRF}}(X) = \int_{\Pi_x} f(x)\log{\left(\frac{f(x)}{f_{GRF}(x)}\right)}dx .
    \label{eqn_turb_interm}
\end{eqnarray} 
Thus, turbulence intermittency of $X=\phi,$ $\epsilon$ or $\Omega$ is defined as the KL divergence of the PDF of $X$ in a turbulent flow field, $f(x)$, from the PDF of $X$ in a Gaussian random velocity field, $f_{GRF}(x)$.
Within this framework, we separate the total intermittency $D_{f||f_{G}}(X)$ into turbulence-induced intermittency $D_{f||f_{GRF}}(X)$ (Eq.~\ref{eqn_turb_interm}) and kinematic intermittency $D_{f_{GRF}||f_{G}}(X)$ (Eq.~\ref{eqn4.5}).
For velocity gradients \cite{yakt18, yadon17} and scale-based velocity differences \cite{gbel}, 
direct comparison with Gaussian distribution yields the turbulence intermittency,
since $f_{GRF}=f_{G}$ and kinematic intermittency vanishes for these quantities.
However, for nonlinear quantities such as $\phi, \epsilon$ and $\Omega$, which possess intrinsic kinematic intermittency, the Gamma distributions $f_{GRF}(x)$ in Eq.~(\ref{eqn4}) provide the appropriate baseline against which turbulence intermittency is measured.
Importantly, these baseline Gamma distributions are identical for any Gaussian velocity field, independent of its variance. 
Consequently, turbulence intermittency of different Reynolds number flows can be evaluated by measuring the deviation from the same baseline PDF $f_{GRF}(x)$.
Hereafter, all subsequent references to intermittency implies turbulence-induced intermittency, unless stated otherwise.

\subsection{Analytical example of intermittency quantification}
\label{sec:Analytical_interm}

\begin{figure*}
    \centering
    \begin{tikzpicture}
    \node[anchor=south west, inner sep=0] (img) at (0,0)
    {\includegraphics[width=1\linewidth]{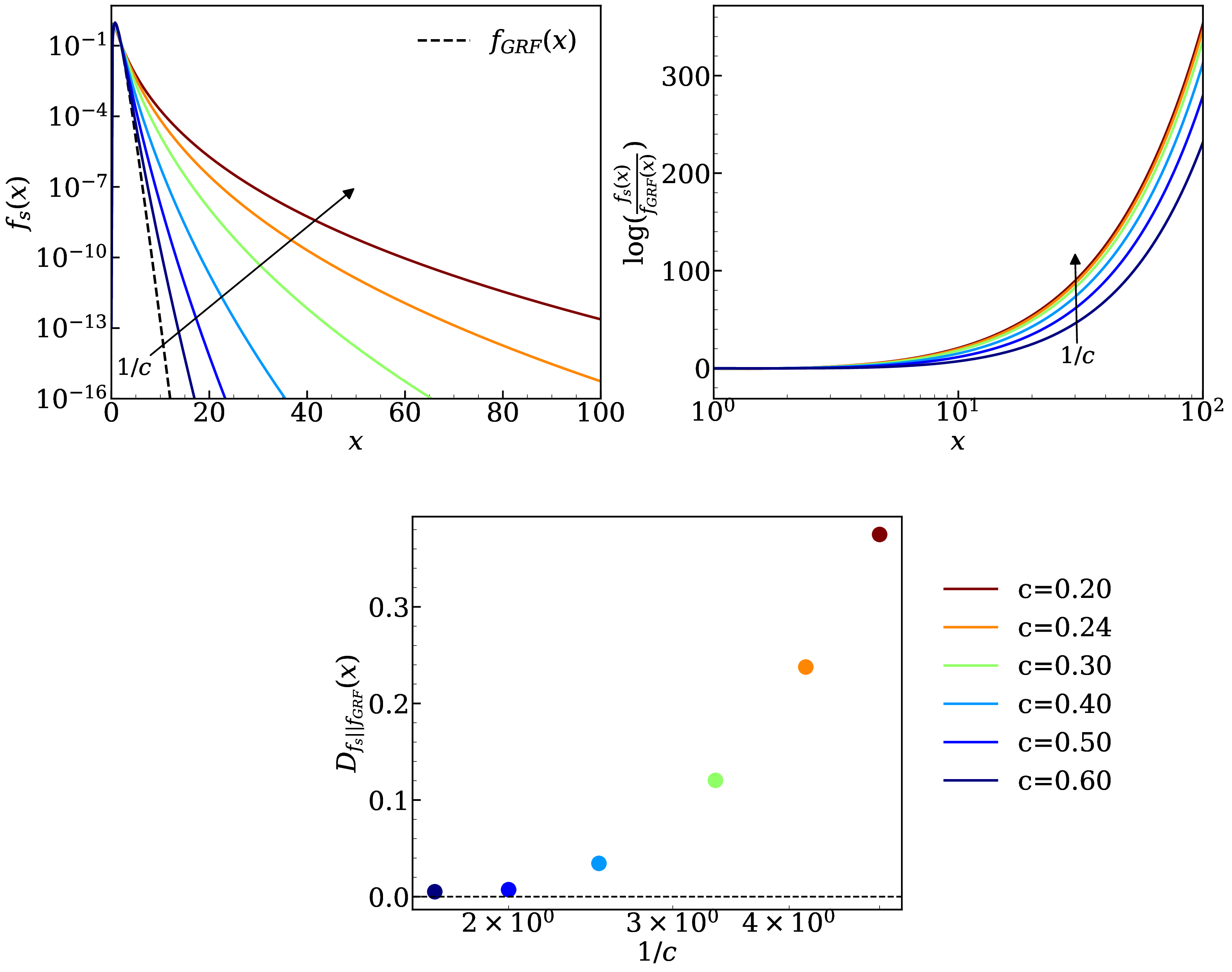}};
        \begin{scope}[x={(img.south east)}, y={(img.north west)}]
              \node at (0.001, 1.0) {\textbf{(a)}};
              \node at (0.505, 1.0) {\textbf{(b)}};
              \node at (0.278, 0.475) {\textbf{(c)}};
        \end{scope}
    \end{tikzpicture}
    
    \caption{Analytical example of a stretched exponential PDF $f_s(x)$: (a) PDFs $f_s(x)$ for different values of exponent $c$ and Gamma PDF $f_{GRF}(\phi)$ for comparison, (b) $\log\left(\frac{f_s(x)}{f_{GRF}(x)}\right)$ - expected value of which yields the KL divergence measure of intermittency $D_{f_s||f_{GRF}}(X)$ and, (c) values of $D_{f_s||f_{GRF}}(X)$ as a function of $1/c$.}
    \label{fig:stretchexpKLD}
\end{figure*}

We present an analytical example representative of small-scale turbulent fields to demonstrate how the proposed KL divergence-based measure quantifies intermittency. To mimic the probability distributions of $\phi$, $\epsilon$ and $\Omega$, we consider the following stretched exponential PDF \cite{yeung2018effects, Gotoh22, kinematic} for a synthetic intermittent field:
\begin{eqnarray}
    f_s(x) = Zx^6\exp (-bx^c)
    \label{eq:stretched}
\end{eqnarray}
where $Z$ is the normalization constant ensuring $\int f_s \, dx=1$.
The exponent $c$ controls the tail heaviness, and is maintained below unity to ensure that the PDF decays with $x$ slower than a pure exponential and exhibits heavier, stretched tails \cite{frisch1995turbulence}.
For a given $c$, the parameter $b$ is prescribed such that the variable has unit mean, similar to the mean-normalized $\phi,\epsilon$ and $\Omega$ studied in this work. 
Fig.~\ref{fig:stretchexpKLD}a illustrates the PDF (Eq.~\ref{eq:stretched}) for different values of $c$. It is evident that decreasing $c$ progressively stretches the tail of the PDF $f_s(x)$, increasing the deviation from the GRF and hence the degree of intermittency.

To understand how the KL divergence $D_{f_s||f_{GRF}}(X)$ captures this, we examine the term $\log\left(\frac{f_s(x)}{f_{GRF}(x)}\right)$, the expected value of which constitutes the intermittency in Eq.~(\ref{eqn_turb_interm}). This term is analogous to $x^{n}$ in the $n^{th}$ order moment conventionally used to study intermittency, but with a key distinction: while $x^{n}$ depends explicitly on the value of $x$, $\log\left(\frac{f_s(x)}{f_{GRF}(x)}\right)$ in KL divergence is purely a function of the PDFs of $x$. As shown in Fig.~\ref{fig:stretchexpKLD}b, this term grows rapidly towards the tail of the PDF suggesting that the KL divergence measure is strongly dominated by contributions from the extreme values.
Finally, the variation of KL divergence with $1/c$ in Fig.~\ref{fig:stretchexpKLD}c shows that 
as the PDF gets more heavy-tailed (lower $c$), the KL divergence value increases, confirming that KL divergence is a robust, suitable measure for quantifying intermittency of turbulent fields. 

This analytical example is also used to examine the sensitivity of KL divergence to sample size and estimation errors, as discussed in detail in Appendix~\ref{appB}. 
The fidelity of the estimation of $D_{f_s||f_{GRF}}(X)$ relies on the accuracy of the estimated PDF from discrete samples.
For a given sample size, there is a specific range of bin width values from which one can accurately estimate KL divergence, free of any estimation bias.

\section{Direct Numerical Simulation (DNS) data}\label{sec:DNS}

To study small-scale intermittency, we analyze DNS data of forced homogeneous isotropic turbulence over a range of $Re_{\lambda} \approx 1$ to $600$ in a periodic $(2\pi)^3$ domain. Details of all the cases are mentioned in Table~\ref{table2}. 
Part of the database is created from simulations performed using an in-house pseudospectral solver \cite{rogallo1981}. 
The solver employs a low-storage second-order Runge–Kutta time integration scheme \cite{canuto2012spectral}, with exact treatment of the viscous term, and a parallel 3D Fast Fourier Transform library \cite{pekurovsky2012p3dfft}. 
To avoid aliasing errors in the solution, the solver uses a phase-shifting method \cite{orszag1972numerical} together with spectral truncation that removes Fourier modes of wavenumber $|\mathrm{\bm{\kappa}}|>\sqrt{2} N/3$ where $N$ is the number of grid points in each direction. 
To maintain a statistically stationary turbulent flow, energy is injected within a low wavenumber shell, $|\mathrm{\bm{\kappa}}| \leq 2$, using a large-scale forcing proportional to the local velocity. This deterministic forcing scheme has been used previously in several DNS studies \cite{machiels1997predictability,ishihara2007small,mccomb2015energy,khurshid2023emergence}.

\begin{table}
\centering
\caption{\label{table2}Description of DNS datasets of forced homogeneous isotropic turbulent flows used in this work: Taylor Reynolds number ($Re_{\lambda}$), number of grid points ($N^3$), resolution given by the maximum wavenumber resolved ($\kappa_{max}$) normalized by Kolmogorov length scale ($\eta$), 
ratio of large-eddy turnover time ($T_L$) and Kolmogorov time scale ($\tau_\eta$), 
number of time snapshots used for averaging statistics ($n_s$),
time duration of averaging in units of large-eddy turnover time ($T_{avg}/T_{L}$), and the source of the data. 
\vspace{0.5em}}
{\renewcommand{\arraystretch}{0.75}
\begin{tabular}{@{\hspace{6pt}}c@{\hspace{14pt}}c@{\hspace{14pt}}c@{\hspace{14pt}}c@{\hspace{14pt}}c@{\hspace{14pt}}c@{\hspace{14pt}}c@{\hspace{6pt}}}
\hline
$Re_{\lambda}$ & $N^3$ & $\kappa_{max}\eta$ & $T_L/\tau_\eta$ & $n_s$ & $T_{avg}/T_{L}$ & Source \\ \hline
1   & $256^3$  & 105.6 & 5.04 & 1 & - & \cite{yadon17} \\
6   & $256^3$  & 34.8  & 5.51 & 1 & - & \cite{yadon17} \\
9   & $256^3$  & 26.6  & 5.67 & 1 & - & \cite{yadon17} \\
14  & $256^3$  & 19.87 & 5.78 & 1 & - & \cite{yadon17} \\
21  & $128^3$  & 7.26  & 5.06 & 217 & 9.9 & in-house solver \\
25  & $256^3$  & 11.51 & 6.48 & 1 & - & \cite{yadon17} \\
37  & $128^3$  & 4.08  & 6.12 & 406 & 48.7 & in-house solver \\
59  & $256^3$  & 4.54  & 7.40 & 542 & 34.7 & in-house solver \\
65  & $256^3$  & 2.91  & 9.49 & 7 & - & in-house solver \\
86  & $256^3$  & 2.83  & 11.57 & 1 & - & \cite{don18} \\
119 & $512^3$  & 3.38  & 13.47 & 1040 & 5.5 & in-house solver \\
252 & $512^3$  & 1.38  & 29.30 & 66 & 15.0 & in-house solver \\
332 & $1024^3$ & 1.5   & 41.92 & 111 & 3.9 & in-house solver \\
385 & $1024^3$ & 1.41  & 44.00 & 1 & - & \cite{don18} \\
427 & $1024^3$ & 1.32  & 46.11 & 1 & - & \cite{Li_08} \\
588 & $2048^3$ & 1.39  & 75.31 & 1 & - & \cite{don18} \\ \hline
\end{tabular}}
\end{table}

Additionally, DNS datasets of similar forced isotropic turbulent flows from the Johns Hopkins Turbulence Database and Donzis research group at Texas A\&M University are also used for the analysis \cite{yadon17, Li_08}. These datasets have been used extensively in the past to study intermittency, scaling of high order moments, and velocity gradient dynamics \cite{yadon17, don18,Das_Girimaji_2019,das2020revisiting, lin2023experimental}. 
All DNS data are at resolutions: $\kappa_{max} \eta > 1.3$, where $\kappa_{max}$ is the maximum resolved wave number, and $\eta$ is the Kolmogorov length scale.

Accurate estimation of the PDFs is essential for reliable computation of information-theoretic measures. 
PDFs of the three quantities ($f(x)$ for $x=\phi,\epsilon,\Omega$) for all DNS datasets are estimated using the histogram method with a constant bin width of $\Delta x = 0.008$. This bin width is selected from a range of optimal $\Delta x$ over which the KL divergence $D_{f||f_{GRF}}(X)$ remains nearly constant, ensuring robust estimation across all $Re_{\lambda}$ (Appendix~\ref{appB}). 
The in-house pseudospectral DNS datasets consist of multiple time instances 
of the statistically stationary flow simulations, conducted over a sufficiently long time duration (see Table~\ref{table2}), while datasets obtained from external sources typically consist of a single or limited time snapshots. 
When multiple time instances are available, the information theoretic measures are averaged across them, and the mean value is reported in the figures throughout the manuscript. For cases with more than $50$ time instances, the error bars based on standard error are also reported about the mean, denoted by hollow (unfilled) symbols in the figures.
The baseline PDF $f_{GRF}(x)$ is evaluated analytically (Eq.~\ref{eqn4}) at the bin centers.

\section{Reynolds number variation of PDF}\label{sec:pseudo}

\begin{figure*}
  \centering
  \begin{tikzpicture}
    \node[anchor=south west, inner sep=0] (img) at (0,0) {\includegraphics[width=\textwidth]{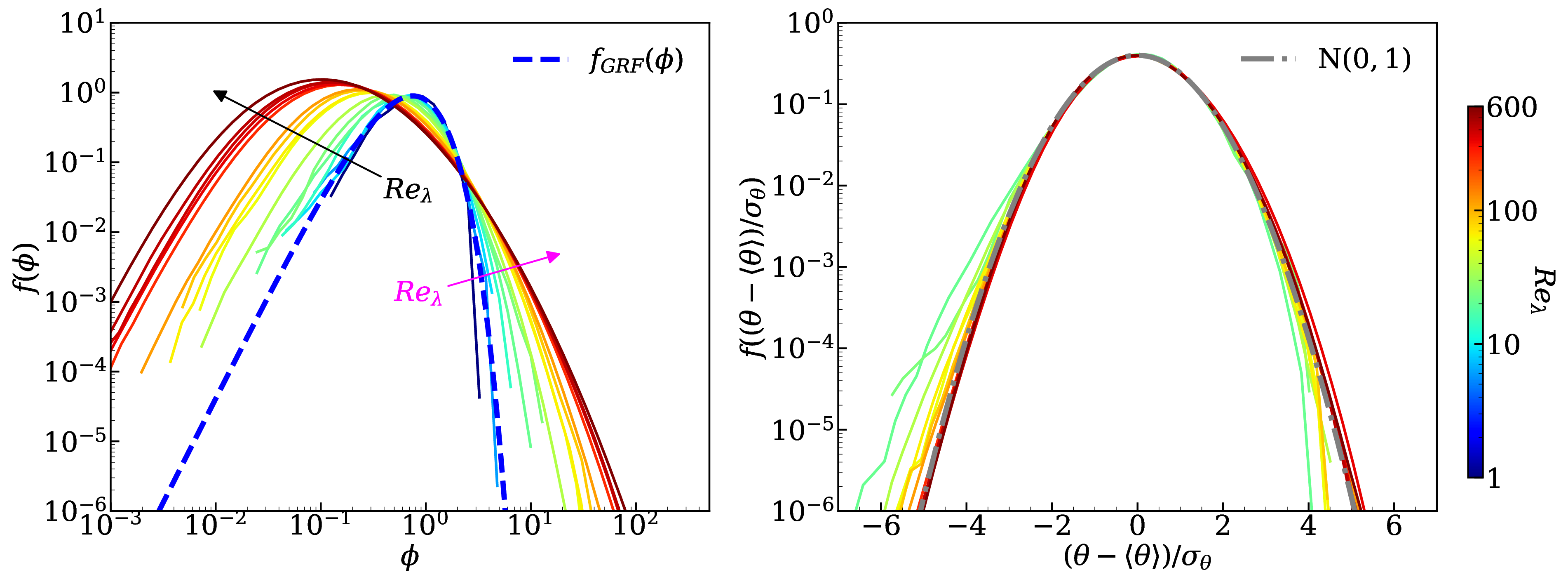}};
    \begin{scope}[x={(img.south east)}, y={(img.north west)}]
      \node at (0.0, 1.0) {\textbf{(a)}};
      \node at (0.47, 1.0) {\textbf{(b)}};
    \end{scope}
  \end{tikzpicture}
  \vspace{-1.5em}
  \caption{\label{fig:3} (a) PDFs of $\phi$ from DNS data of all $Re_{\lambda}$, $f(\phi)$, and from baseline Gaussian field, $f_{GRF}(\phi)$ (blue dashed line); arrows indicate the direction of increasing $Re_{\lambda}$. 
  (b) PDFs of standard normalized $\theta (\equiv \log(A_{ij}A_{ij}) = \log(\phi) + \log\langle A_{ij}A_{ij}\rangle$) from DNS data of  $Re_{\lambda}>20$ and standard normal distribution $\mathcal{N}(0,1)$ (grey dashed-dotted line).}
\end{figure*}

We examine the probability distribution of pseudodissipation rate ($ \phi = A_{ij}A_{ij} / \langle A_{ij}A_{ij}\rangle$) in detail, as it represents the total magnitude of the velocity gradient tensor, with combined contributions of both dissipation rate and enstrophy.
Fig.~\ref{fig:3}a illustrates the PDFs $f(\phi)$ for isotropic turbulent flows of different Reynolds numbers. It is evident that at very low $Re_{\lambda}$, the PDF is nearly identical to that of a GRF.
As $Re_{\lambda}$ increases, the tail of the PDF grows progressively heavier, signifying an increasing departure from the non-intermittent GRF baseline, indicated by magenta arrow in Fig.~\ref{fig:3}a. The peak of the PDF simultaneously shifts toward lower values of $\phi$, suggesting an increasing density of regions of extremely low velocity gradients in the flow, indicated by black arrow in Fig.~\ref{fig:3}a. Similar trends are observed for dissipation and enstrophy PDFs, as reported previously \cite{kinematic, Gotoh22, Yeung_Donzis_Sreenivasan_2012}.

Prior studies have shown that at high enough Reynolds numbers, $A_{ij}A_{ij}$ follows a near-lognormal distribution \cite{Das_Girimaji_2024, Yeung_Pope_1989}, that is, $\theta\equiv \log{(A_{ij}A_{ij})}$ is approximately Gaussian. This is evident from the PDFs plotted in Fig.~\ref{fig:3}b, particularly for higher $Re_{\lambda}$.
Thus, the lognormal (LN) model for $\phi \;(= \exp(\theta)/\langle A_{ij}A_{ij}\rangle)$ can be derived as follows:
\begin{equation}
    f_{LN}(\phi) = \frac{1}{\sqrt{2\pi}\sigma_{\theta}\phi} \exp\left[-\frac{1}{2\sigma_{\theta}^2}\left(\log{\left(\phi\right)} + \frac{1}{2}\sigma_{\theta}^2\right)^2\right].
    \label{eqn8}
\end{equation}
For $Re_{\lambda}>20$, this provides a reasonable approximation of the PDF over moderate values of $\phi$, which constitute the bulk of the flow field.
The model $f_{LN}(\phi)$ depends only on the variance of $\theta$ $(\sigma_{\theta}^2)$, which increases logarithmically with $Re_{\lambda}$ following $\sigma_{\theta}^2 \approx -0.354 + 0.289\log(Re_\lambda)$ \cite{Yeung_Pope_1989, Das_Girimaji_2024}.
In the subsequent sections, we assess how faithfully the lognormal model PDF $f_{LN}(\phi)$ reproduces the uncertainty and intermittency of $\phi$ against DNS data.

\section{Uncertainty}\label{sec:entropy_results}

\begin{figure}
\includegraphics[width=0.6\columnwidth]{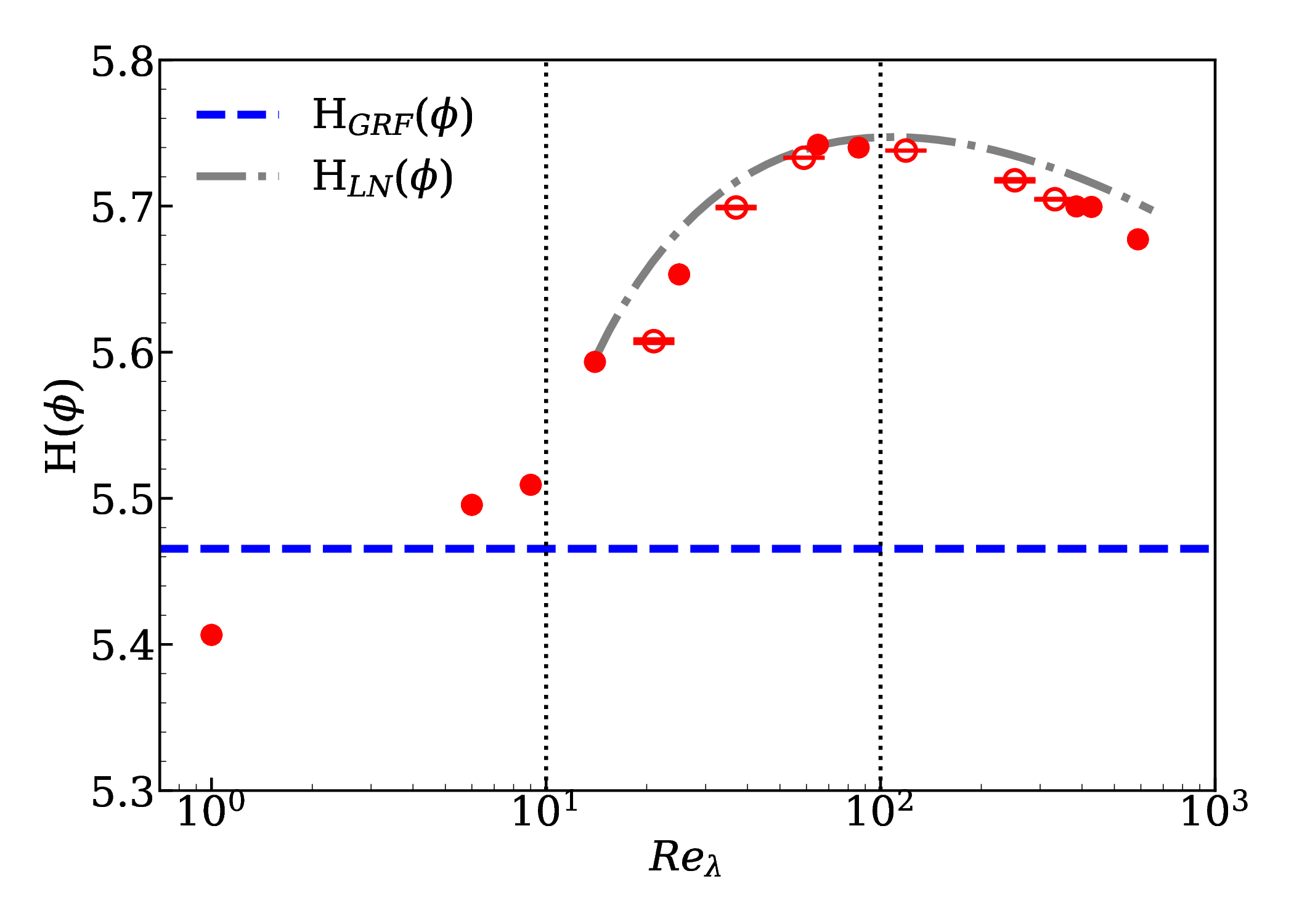}
\vspace{-1.5em}
\caption{\label{fig:4} Shannon entropy of pseudodissipation rate, $\mathrm{H}\left(\phi\right)$, from DNS data of different $Re_{\lambda}$. Blue dashed line: entropy in GRF, $\mathrm{H}_{GRF}\left(\phi\right)\approx 0.637 - \log(\Delta \phi)$, derived from Eqs.~(\ref{eqn3},\ref{eqn4}) \cite{cover} and grey dashed-dotted line: entropy based on lognormal fit, $\mathrm{H}_{LN}\left(\phi\right) = \log{(\sqrt{2\pi e} /\Delta \phi)} + \log{(\sigma_{\theta}^2)}/2 - \sigma_{\theta}^2/2$.
Black dotted lines denote the transitional Reynolds numbers marking a change in regime.}
\end{figure} 

Shannon entropy or uncertainty (Eq.~\ref{eqn3}) of pseudodissipation rate, $\mathrm{H}\left(\phi\right)$, is estimated from DNS data and plotted as a function of $Re_{\lambda}$ in Fig.~\ref{fig:4}. 
For comparison, the figure also shows the entropy of $\phi$ in a non-intermittent Gaussian random velocity field (Eq.~\ref{eqn4}) \cite{cover},
\begin{equation}
    \mathrm{H}_{GRF}\left(\phi\right) \approx 0.637 - \log(\Delta \phi)
\end{equation}
and the entropy given by the lognormal model (Eq.\ref{eqn8}),
\begin{equation}
    \mathrm{H}_{LN}(\phi)=\frac{1}{2}\log{(2\pi e)} + \frac{1}{2}\log{(\sigma_{\theta}^2)} - \frac{\sigma_{\theta}^2}{2} - \log{(\Delta \phi)},
    \label{eq:HLN}
\end{equation}
which shows a reasonable approximation of the entropy from the DNS data at higher {$Re_{\lambda}$}.

The variation of $H(\phi)$ in a turbulent flow can be divided into three regimes. At low Reynolds numbers ($Re_{\lambda}\approx 0$ to $10$), the entropy is close to $H_{GRF}$ with a considerable scatter, indicating that 
the flow is nearly Gaussian at such low Reynolds numbers.
At moderate Reynolds numbers ($Re_{\lambda}\approx 10$ to $100$), the entropy grows rapidly with $Re_{\lambda}$, suggesting a growth in uncertainty of pseudodissipation rate as turbulence develops.
Note that this change of behavior around $Re_{\lambda}\approx 10$ is consistent with previous observations of Yakhot and Donzis \cite{yadon17, yakt18} who showed a rapid deviation of velocity gradient moments from Gaussian values beyond a similar transition $Re_{\lambda}$.
Surprisingly, at high Reynolds numbers ($Re_{\lambda}> 100$), the uncertainty associated with the pseudodissipation field declines with further increase in Reynolds number.

The observed variation of $\mathrm{H}(\phi)$ arises from two competing effects in the PDF of $\phi$ (Fig.~\ref{fig:3}a): entropy decreases as outcomes become less equiprobable, reflected in narrowing and heightening of the PDF peak (black arrow), and entropy increases as more outcomes become possible, reflected in the stretching of the PDF tails (magenta arrow) \cite{cover, Gotoh22, kinematic}. 
In the range $10 \leq Re_{\lambda} \leq 100$, the latter effect dominates -- tail stretching drives the net increase in entropy, signifying the increasingly non-Gaussian nature of the flow as turbulence develops.
However, above $Re_{\lambda} \approx 100$, the former effect dominates -- entropy declines as the PDF deforms toward low-$\phi$ events, developing a sharper peak.
This implies that a narrow range of very low pseudodissipation events begins to dominate strongly over most of the flow domain, and despite the increase in extreme events, the net result is a decline in entropy $H(\phi)$ with increasing $Re_{\lambda}$.
The peak of $H(\phi)$ near $Re_{\lambda}\approx 100$ is further corroborated by the lognormal PDF model wherein $\mathrm{H}_{LN}(\phi)$ (Eq.~\ref{eq:HLN}) attains its maximum at $\sigma_{\theta}^2 = 1$, corresponding to $Re_{\lambda} = 108$.
This suggests that the observed non-monotonic trend of entropy is a genuine feature of the turbulence pseudodissipation rate, supported by the analytically calculated $\mathrm{H}_{LN}(\phi)$, rather than an artifact of statistical estimation.

Interestingly, the Reynolds number of peak entropy ($Re_{\lambda} \approx 100$) nearly coincides with the convergence of several key small-scale statistics: normalized mean dissipation (dissipative anomaly) \cite{sreenivasan1998update, donzis2005scalar}, average velocity-gradient partitioning into shear, normal-strain and rigid-body-rotation \cite{das2020revisiting,arun2024velocity}, and the joint PDF of normalized second and third velocity-gradient invariants ($q,r$), representing the local streamline shape \cite{das2022effect}.
These suggest that this $Re_{\lambda}$ potentially marks the onset of fully developed turbulence, above which any further increase in Reynolds number paradoxically leads to a more ``ordered" (lower uncertainty) state of small-scale turbulence: a state where more of the flow domain is dominated by a progressively narrowing range of extremely low-pseudodissipation events. 

\begin{figure}
    \centering
    \includegraphics[width=0.6\linewidth]{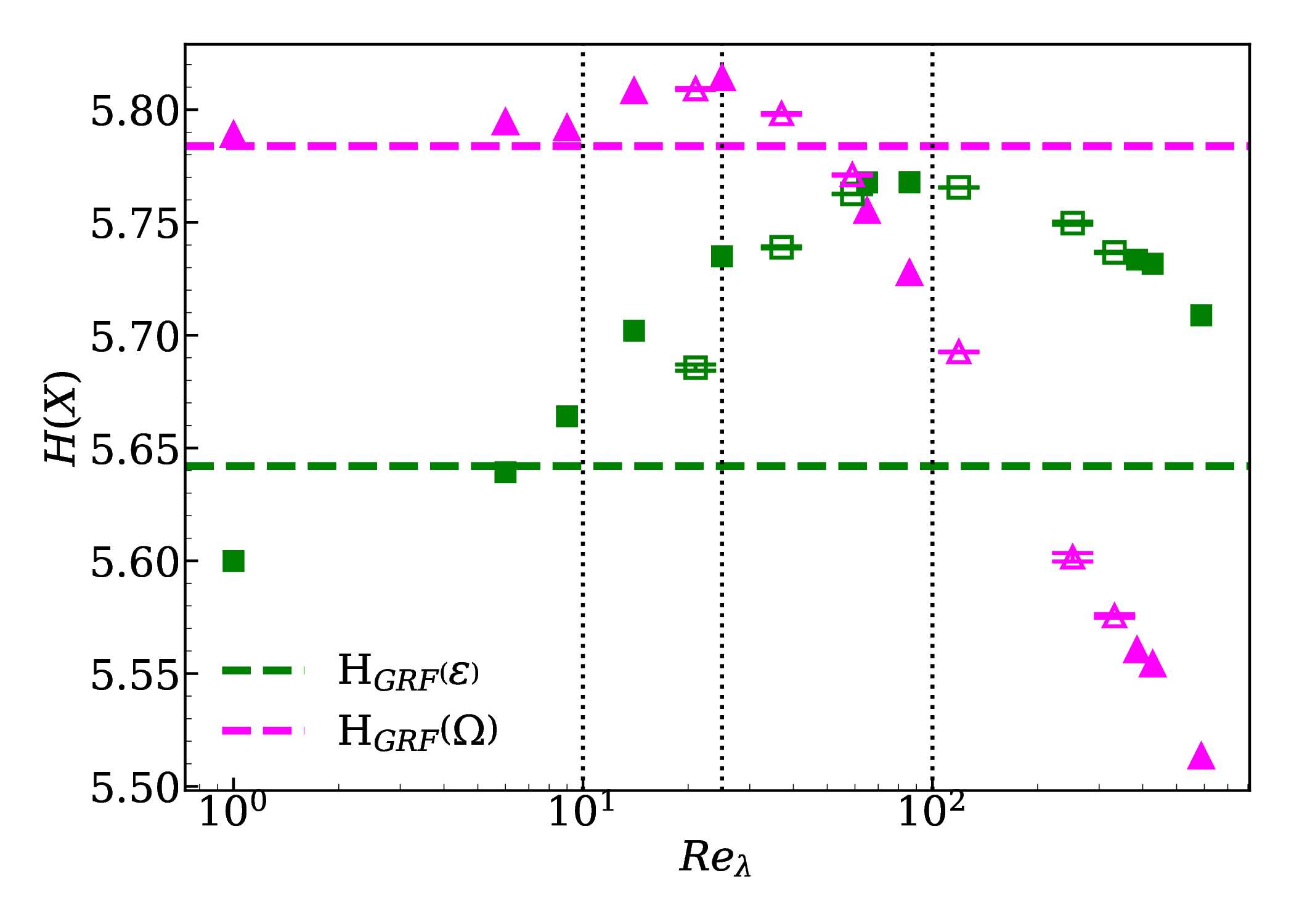}
    \vspace{-1.5em}
    \caption{Shannon entropy of dissipation rate, $\mathrm{H}\left(\epsilon\right)$ in green and of enstrophy, $\mathrm{H}\left(\Omega\right)$ in magenta, from DNS data of different $Re_{\lambda}$. Green dashed line: entropy of dissipation in GRF, $\mathrm{H}_{GRF}\left(\epsilon\right)\approx 0.814 - \log(\Delta \epsilon)$ and magenta dashed line: entropy of enstrophy in GRF, $\mathrm{H}_{GRF}\left(\Omega\right)\approx 0.956 - \log(\Delta \Omega)$, derived from Eqs.~(\ref{eqn3},\ref{eqn4}) \cite{cover}.
Black dotted lines denote the transitional Reynolds numbers marking a change in regime.}
    \label{fig:H_eps_ome}
\end{figure}

Shannon entropies of dissipation rate and enstrophy as a function of Taylor Reynolds number are plotted in Fig.~\ref{fig:H_eps_ome}, along with the 
corresponding entropy values for the baseline GRF field (Eq.~\ref{eqn4}):
\begin{eqnarray}
    \mathrm{H}_{GRF}\left(\epsilon \right) \approx {0.814} - \log(\Delta \epsilon), \;\; 
    \mathrm{H}_{GRF}\left(\Omega \right) \approx {0.956} - \log(\Delta \Omega).
\end{eqnarray}
Similar to $\phi$ (Fig.~\ref{fig:4}), the uncertainty of both $\epsilon$ and $\Omega$ fields also exhibit a three-regime variation with $Re_{\lambda}$.
At low Reynolds numbers ($Re_{\lambda}<10$), both $H(\epsilon)$ and $H(\Omega)$ are close to the corresponding GRF values, reinforcing that the flow is nearly Gaussian in this regime \cite{khurshid2023emergence, yadon17, yakt18, yakhot2017multiscaling}.
Above the transitional $Re_{\lambda} \approx 10$, the entropy of both fields rises as turbulence develops, $\epsilon$ more rapidly than $\Omega$.
For dissipation rate, this growth in uncertainty persists until $Re_{\lambda} \approx 100$, beyond which it declines, demonstrating a trend qualitatively consistent with that of pseudodissipation rate.
Enstrophy, however, begins to decline at a much lower $Re_{\lambda} \approx 25$ and falls well below its GRF value at high Reynolds numbers. As discussed above, these entropy variations in both fields stem from the balance between the PDF's peak steepening and tail stretching.
The $\Omega$ field undergoes a more rapid steepening of the PDF peak at low values with increasing $Re_{\lambda}$ \cite{kinematic}, the entropy reduction due to which outweighs the increase due to intensification of extreme events, at a considerably lower $Re_{\lambda}$ than $\epsilon$ or $\phi$.
The results indicate that low enstrophy events become concentrated within a progressively contracting range of values at a much earlier stage of turbulence development compared to dissipation or pseudodissipation, ultimately resulting in an enstrophy field which is significantly lower in entropy than the other two fields in fully developed turbulence of high $Re_{\lambda}$.

\section{Intermittency}\label{sec:interm_results}

\begin{figure}
\begin{tikzpicture}
\node[anchor=south west, inner sep=0] (img) at (0,0)
{\includegraphics[width=\columnwidth]{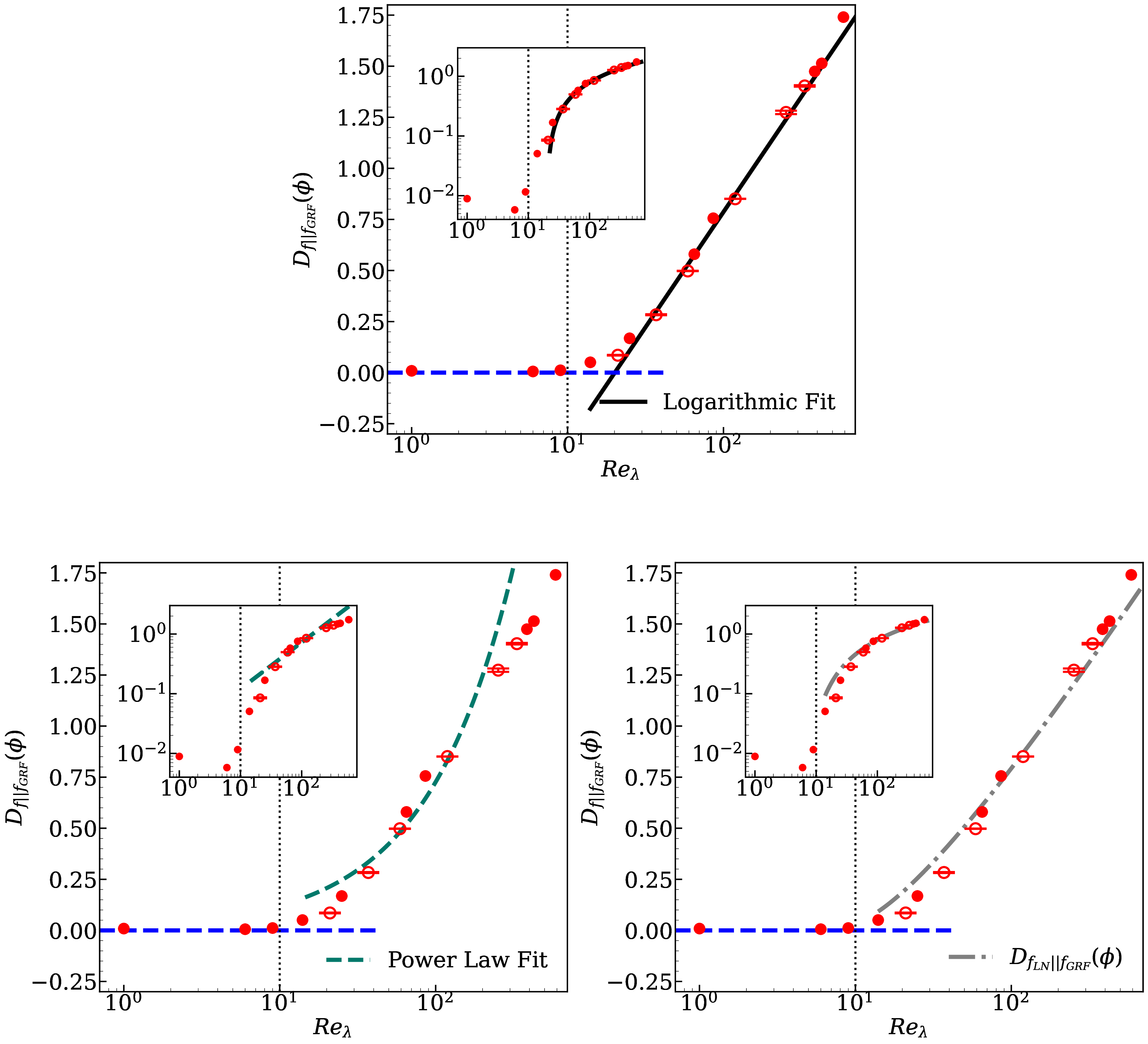}};
    \begin{scope}[x={(img.south east)}, y={(img.north west)}]
          \node at (0.25, 1.0) {\textbf{(a)}};
          \node at (0.0, 0.48) {\textbf{(b)}};
          \node at (0.52, 0.48) {\textbf{(c)}};
    \end{scope}
\end{tikzpicture}
\caption{\label{fig:5} KL divergence of turbulence pseudodissipation rate from GRF,  
$D_{f||f_{GRF}}(\phi)$, as a function of $Re_{\lambda}$ in log-linear scale (log-log scale in inset). 
Blue dashed lines indicate constant fit $D_{f||f_{GRF}}(\phi) \approx 0$ for $Re_{\lambda} \leq 10$. 
Following fits for $Re_{\lambda} \in (20,600)$:
(a) log-law fit ($R^2$ = 0.996) $D_{f||f_{GRF}}(\phi) \approx -1.47 +  0.49\log{\left(Re_{\lambda}\right)}$. 
(b) power law fit ($R^2$ = 0.899)  $D_{f||f_{GRF}}(\phi) \approx 0.015Re_{\lambda}^{0.784}$.
(c) lognormal model 
$D_{f_{LN}||f_{GRF}}(\phi) = \log{\left(3e^{7/2}/128\sqrt{2\pi}\right)} + 2\sigma_{\theta}^2 - \frac{1}{2}\log{\sigma_{\theta}^2}$.}
\end{figure} 

Previous works have demonstrated that the higher-order moments of dissipation rate and enstrophy exhibit power-law scaling $\sim Re_{\lambda}^{\beta_{n}}$, where the scaling exponent ${\beta_{n}}$ varies with the moment order $n$ \cite{don18, zhou2000reynolds, vela2022energy, khurshid2023emergence, buaria2022scaling, Elsinga_Ishihara_Hunt_2023}.
Quantifying intermittency using KL divergence $D_{f||f_{GRF}}$ integrates the effects of all these moments in one measure, and isolates the intermittency arising solely from turbulence dynamics separating the kinematic contribution, as discussed in section~\ref{sec:interm}.
The KL divergence of pseudodissipation rate, $D_{f||f_{GRF}}(\phi)$, is plotted as a function of $Re_{\lambda}$ in Fig.~\ref{fig:5}. 
At very low Reynolds numbers ($Re_{\lambda} \leq 10$), pseudodissipation rate exhibits nearly zero turbulence intermittency and is close to a Gaussian random velocity field. 
The figure clearly illustrates a transitional $Re_{\lambda} \approx 10$, above which $D_{f||f_{GRF}}(\phi)$ grows rapidly with Reynolds number, reflecting the field's deviation from GRF and increasing intermittency.
This transitional behavior is consistent with that previously observed in the moments of longitudinal velocity gradients \cite{yakt18,Das_Girimaji_2019}.
Beyond this transition, $D_{f||f_{GRF}}(\phi)$ increases logarithmically with $Re_{\lambda}$:
\begin{eqnarray}
    D_{f||f_{GRF}}(\phi) \approx -1.47 + 0.49\log (Re_{\lambda})
    \label{eqn9.75}
\end{eqnarray}  
in the range $Re_{\lambda}\in (20,600)$.
Note that there is a short range of transition, $Re_{\lambda} \in (10,20)$, that separates the near-zero intermittency (Gaussian) regime from the onset of logarithmic growth.
The logarithmic growth of KL divergence indicates that the turbulence intermittency of $\phi$ increases with $Re_{\lambda}$ but at a much slower rate than that suggested by the power law scaling of its moments \cite{yadon17, yakt18}. The growth rate particularly slows down at high $Re_{\lambda}$ as evident in the log-log scale in the inset of Fig.~\ref{fig:5}a.
Note that the error bars, based on the standard error across multiple time snapshots of a simulation, are negligibly small, suggesting that the variability of KL divergence is significantly lower than that typically observed for higher-order moments.

For comparison, the best-fit power law for $D_{f||f_{GRF}}(\phi)$ is illustrated in Fig.~\ref{fig:5}b. It is evident in both linear-log and log-log scale that the growth of turbulence intermittency for $Re_{\lambda}>10$ does not follow a power law scaling, and is significantly better captured by a logarithmic scaling.
The lognormal model (Eq.~\ref{eqn8}), on the other hand, yields the following KL divergence:
\begin{equation}
    D_{f_{LN}||f_{GRF}}(\phi) \sim 2\sigma_{\theta}^2 - \frac{1}{2}\log{\sigma_{\theta}^2},
\end{equation}
which shows a reasonable approximation of this growth (Fig.~\ref{fig:5}c), though less accurately than the logarithmic fit (Eq.~\ref{eqn9.75}).
Since $\sigma_{\theta}^2 \sim \log{Re_{\lambda}}$, in the limit of $Re_{\lambda} \rightarrow \infty$, we get $\log(\sigma_{\theta}^2) \ll \sigma_{\theta}^2$, and consequently $D_{f_{LN}||f_{GRF}}(\phi) \sim \log(Re_{\lambda})$. Thus, at sufficiently high Reynolds numbers, the lognormal model recovers a logarithmic growth of turbulence intermittency.
Overall, it is important to note that while there is evidence of structural convergence and decline of entropy of small scales beyond $Re_{\lambda}>100$, the small-scale intermittency induced by turbulence dynamics continues to grow monotonically following the same logarithmic scaling. 

\begin{figure}
\includegraphics[width=0.6\columnwidth]{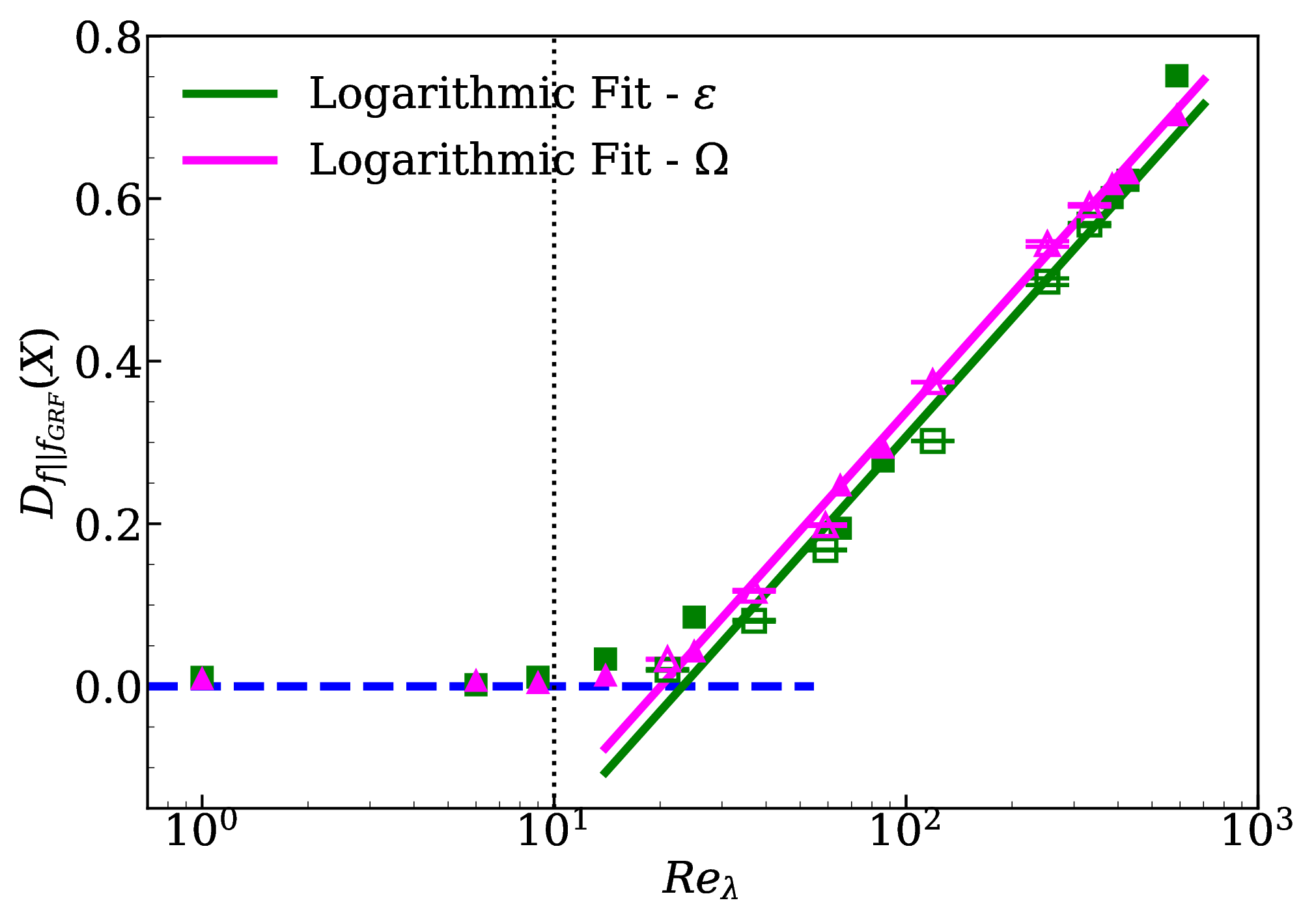}
\vspace{-1.8em}
\caption{\label{fig:6} KL divergence of dissipation rate $D_{f||f_{GRF}}(\epsilon)$ (green squares) and enstrophy $D_{f||f_{GRF}}(\Omega)$ (magenta triangles) as a function of $Re_{\lambda}$. Blue dashed line indicates constant fit  $D_{f||f_{GRF}}(X) \approx 0$ for $Re_{\lambda} \leq 10$, and solid lines indicate the log-law fits $D_{f||f_{GRF}}(\epsilon)\approx -0.66 + 0.21\log{\left(Re_{\lambda}\right)}$ ($R^2$ = 0.980) and $D_{f||f_{GRF}}(\Omega)\approx -0.63 + 0.21\log{\left(Re_{\lambda}\right)}$ ($R^2$ = 0.997) for $Re_{\lambda} \in (20,600)$.}
\end{figure}

KL divergence of dissipation rate $\epsilon$ and enstrophy $\Omega$ follow a similar trend with Reynolds number as $\phi$ (Fig.~\ref{fig:6}). 
Below the transitional $Re_{\lambda}\approx 10$,
both fields have nearly zero turbulence intermittency and exhibit the Gamma distribution (Eq.~\ref{eqn4}) of the baseline GRF in agreement with the findings of Gotoh et al. \cite{kinematic}.
Above this $Re_{\lambda}$, both exhibit logarithmic growth in KL divergence (see Appendix~\ref{appC} for a comparison with the corresponding power-law fits). The approximate logarithmic scaling of $\epsilon$ and $\Omega$ are given by:
\begin{subequations}
\label{eqn11}
\begin{eqnarray}
    D_{f||f_{GRF}}(\epsilon) \approx -0.66 + 0.21\log{(Re_{\lambda})} \label{eqn11a} \\
    D_{f||f_{GRF}}(\Omega) \approx -0.63 + 0.21\log{(Re_{\lambda})}. \label{eqn11b}
\end{eqnarray}
\end{subequations}
Surprisingly, the intermittency of $\epsilon$ and $\Omega$ relative to their respective baseline GRF distributions, are nearly identical to each other across all Reynolds numbers.
This is counterintuitive, as $\Omega$ has traditionally been considered more intermittent than $\epsilon$ due to its elevated higher-order moments and heavier-tailed PDF \cite{siggia1981numerical, kerr1985higher,buaria2019extreme,Elsinga_Ishihara_Hunt_2023}.
These differences are a consequence of kinematic intermittency \cite{tsinober, tsinober2019essence, chen1995kolmogorov, chen1998kinematic}, present even in non-intermittent Gaussian random velocity fields, where $f_{GRF}(\Omega)$ inherently has a heavier tail than $f_{GRF}(\epsilon)$ (Fig.~\ref{fig:1}). Thus, our results suggest that turbulence dynamics induces nearly equal intermittency in dissipation and enstrophy (Fig.~\ref{fig:6}).
Although unexpected, 
certain prior observations in literature indicate a possibility of this finding:
at high $Re_{\lambda}$, $\epsilon$ and $\Omega$ have nearly equal PDF-stretching exponents \cite{kinematic, don18}, similar PDF shapes \cite{yeung2018effects}, comparable scaling exponents of moments \cite{Elsinga_Ishihara_Hunt_2023}, and Reynolds number collapse of PDFs when scaled by the same time scale \cite{buaria2022vorticity}. 
Comparing Figs.~\ref{fig:5} and \ref{fig:6}, we further observe that pseudodissipation, despite being the least kinematically intermittent in a Gaussian random field (Fig.~\ref{fig:1}), exhibits greater intermittency due to turbulence dynamics than dissipation or enstrophy, and its intermittency grows more rapidly with Reynolds number. 

\section{Conclusion}

This work characterizes the small-scale intermittency and uncertainty in turbulence by studying KL divergence and Shannon entropy of pseudodissipation, dissipation and enstrophy.
KL divergence of these turbulent fields relative to a Gaussian random velocity field quantifies intermittency arising purely from turbulence dynamics, isolating it from the kinematic contribution \cite{kinematic, tsinober, tsinober2019essence}.
Additionally, KL divergence provides a more complete characterization of intermittency based on the probability distribution as a whole rather than individual moments.
Analysis of DNS data of forced homogeneous isotropic turbulence shows that at very low Taylor Reynolds numbers, the flow field is close to a Gaussian random velocity field with nearly zero turbulence intermittency, and above the transitional Reynolds number, $Re_{\lambda} \approx 10$, the field begins to deviate from Gaussianity as the turbulence intermittency grows monotonically.
Contrary to the commonly observed power-law scaling, turbulence intermittency is shown to scale logarithmically with $Re_{\lambda}$.
This suggests that Navier-Stokes dynamics causes the growth rate of intermittency to decelerate at higher Reynolds numbers, asymptotically converging to zero in the limit of infinitely large Reynolds number. 
Remarkably, we find that dissipation and enstrophy exhibit nearly equal turbulence intermittency at all Reynolds numbers, 
revealing a symmetry between strain-rate and vorticity in turbulence dynamics. 
The statistical asymmetry commonly inferred from the PDFs and moments of enstrophy and dissipation 
\cite{Yeung_Donzis_Sreenivasan_2012, buaria2022vorticity, buaria2022scaling} does not stem from turbulence dynamics but rather from their kinematic definitions 
as evidenced by the persistence of the same asymmetry even in a non-intermittent uncorrelated Gaussian random velocity field.

The Shannon entropy of pseudodissipation, dissipation and enstrophy, an indicator of uncertainty of turbulence small scales, exhibits a non-monotonic variation with Reynolds number. 
Consistent with previous observations, the entropy values indicate near-Gaussian flow behavior at very low Reynolds numbers up to $Re_{\lambda} \approx 10$.
Above this, at moderate Reynolds numbers, the entropy increases reflecting progressive symmetry breaking and increasing uncertainty at small scales as turbulence develops \cite{frisch1995turbulence, bormashenko2019symmetry, bormashenko2019entropy, bormashenko2020entropy}. 
Interestingly, above a certain Reynolds number, the entropy decreases despite the increasing variability and intensification of extreme events. This decay suggests a restoration of statistical symmetry at high Reynolds numbers leading to a less uncertain state of small-scale turbulence \cite{warhaft2002turbulence,antonov2018statistical} wherein the bulk of the flow field exhibits a progressively contracting range of very low pseudodissipation (or dissipation/enstrophy) values.
The peak uncertainty occurs at $Re_{\lambda} \approx 100$ for pseudodissipation and dissipation, but at a considerably lower $Re_{\lambda} \approx 25$ for enstrophy, reflecting a more rapid steepening of the enstrophy PDF peak compared to the other fields.

At sufficiently high Reynolds numbers, a notable divergence emerges: turbulence intermittency continues to rise logarithmically, while Shannon entropy begins to decline. This reflects the widening separation between low-intensity and extreme events — the flow simultaneously concentrates quiescent events within a narrowing range while generating increasingly intense events. Entropy captures this bifurcation in the probability distribution but intermittency does not, demonstrating the complementary inferences from these two information-theoretic quantities.
Notably, Shannon entropy indicates that the distributions of the enstrophy and dissipation fields evolve differently with increasing Reynolds number, their nearly identical intermittency growth suggests a similar amplification of extreme occurrences due to turbulence. 
Overall, this study offers a new quantitative understanding of the statistical structure and probability distributions of small-scale turbulent fields, with broader implications for turbulence modeling and development of a comprehensive statistical theory of turbulence.

\section{Acknowledgments}

The authors gratefully acknowledge the financial support from the United States Army Research Laboratory ITC-IPAC under contract no. FA520924C0025 and the Indian Institute of Science under Startup Research Grant of Rishita Das. 
The authors would like to thank Donzis research group at Texas A\&M University and Johns Hopkins Turbulence Database for providing part of the DNS data, and Mr. Aditya Prajapati and Mr. A. Bhargav Manohar for assistance with the DNS solver. Numerical computations and data processing were performed using the IISc Supercomputer Education and Research Centre (SERC) resources.

\appendix

\section{Derivation of PDFs in a Gaussian random velocity field}
\label{appA}

Consider a zero-mean, divergence-free, and spatially-uncorrelated isotropic Gaussian velocity field. In such a field, each component of the velocity gradient tensor (VGT) $(A_{ij})$ is distributed according to a Gaussian PDF as follows,
\begin{align}
    A_{ij} \sim
    \begin{cases}
    \mathcal{N}(0, \sigma_L^2) & \text{for } i = j \\
    \mathcal{N}(0, \sigma_T^2) & \text{for } i \neq j
    \end{cases}
    \label{eq:AppA1}
\end{align}
where the transverse variances $(\sigma_T^2)$ and the longitudinal variances $(\sigma_L^2)$ are related as $\sigma_T^2/\sigma_L^2 = 2$ from isotropy relations \cite{pope2001turbulent, vreman2014statistics}. Imposing the incompressibility constraint: $A_{33} = - A_{11} - A_{22}$, the expression for the field $A_{ij}A_{ij}$ can be expanded as,
\begin{align}
    A_{ij}A_{ij} 
 &= (2A_{11}A_{22} + 2A_{11}^2 + 2A_{22}^2) + (A_{12}^2 + A_{21}^2 + A_{13}^2 + A_{31}^2 + A_{23}^2 + A_{32}^2)
\end{align}
Here, the first bracket contain all the quadratic terms associated with longitudinal velocity gradients ($A_{ij} \; \forall \; i=j$), which are Gaussian fields but mutually correlated due to the incompressibility constraint. 
The second bracket consists of quadratic terms of transverse velocity gradients ($A_{ij} \; \forall \; i \neq j$), which are uncorrelated Gaussian fields. 
Thus, the sum of the longitudinal terms follows $(2A_{11}A_{22} + 2A_{11}^2 + 2A_{22}^2) \sim \varGamma(1, 4\sigma_L^2) \sim \varGamma(1,2 \sigma_T^2)$, while each of the quadratic terms of transverse velocity gradients individually follows $A_{ij}^2 \sim \varGamma(\frac{1}{2}, 2\sigma_T^2) \; \forall \; i \neq j$, where $\varGamma(a,b)$ represents a random variable following a Gamma distribution with shape parameter $a$ and scale parameter $b$ \cite{Gotoh22}.
Therefore, $A_{ij}A_{ij}$ follows:
\begin{equation}
    A_{ij}A_{ij} \sim \varGamma(1, 2\sigma_T^2) + \sum_{k=1}^{6}\varGamma(\frac{1}{2}, 2\sigma_T^2) \sim \varGamma(\frac{8}{2}, 2\sigma_T^2),
\end{equation}
from which we obtain the following PDF of $A_{ij}A_{ij}$:
\begin{equation}
    f_{GRF}(x) = \frac{1}{\Gamma(4)(2\sigma_T^2)^4}x^3\exp(-\frac{x}{2\sigma_T^2}) \quad \text{where, } \quad x = A_{ij}A_{ij}.
\end{equation}
The mean from this distribution is $\langle A_{ij}A_{ij}\rangle= 8\sigma_T^2$.
Thus, the PDF of mean-normalized random variable $\phi = A_{ij}A_{ij}/\langle A_{ij}A_{ij}\rangle$ in a GRF is given by,  
\begin{align}
    f_{GRF}(\phi) = \frac{4^4}{\Gamma(4)}\phi^3\exp(-4\phi)
\end{align}
Clearly, $\phi \sim \varGamma(\frac{8}{2}, \frac{2}{8})$, where the parameters of this Gamma distribution depend on the number of independent terms in the expansion. The PDFs of mean-normalized dissipation, $\epsilon = S_{ij}S_{ij}/\langle S_{ij}S_{ij} \rangle$, and enstrophy, $\Omega = W_{ij}W_{ij}/\langle W_{ij}W_{ij} \rangle$, can be derived similarly and they follow the Gamma distributions, $\epsilon \sim \varGamma(\frac{5}{2}, \frac{2}{5})$ and $\Omega \sim \varGamma(\frac{3}{2}, \frac{2}{3})$, respectively  (see detailed derivation in \cite{Gotoh22}).

\section{Sensitivity of KL divergence to bin width and sample size} \label{appB}

\begin{figure}
    \centering
    \includegraphics[width=0.7\linewidth]{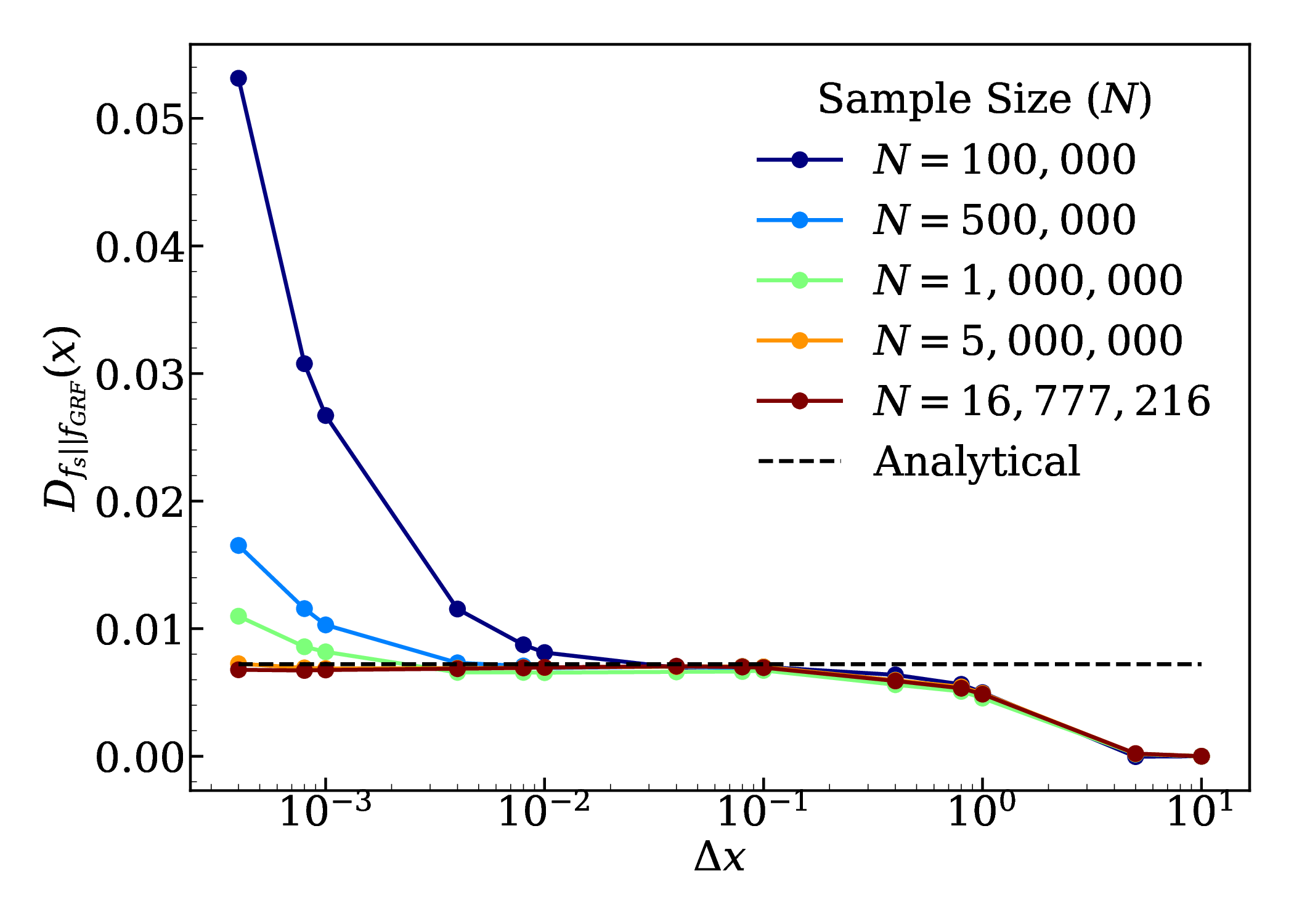}
    \caption{Variation of KL divergence $D_{f_s||f_{GRF}}(x)$ with bin width ($\Delta x$) and sample size ($N$).}
    \label{fig:KLD_sensitivity_N_delx_gamma}
\end{figure}

\begin{figure}
    \centering
    \includegraphics[width=0.7\linewidth]{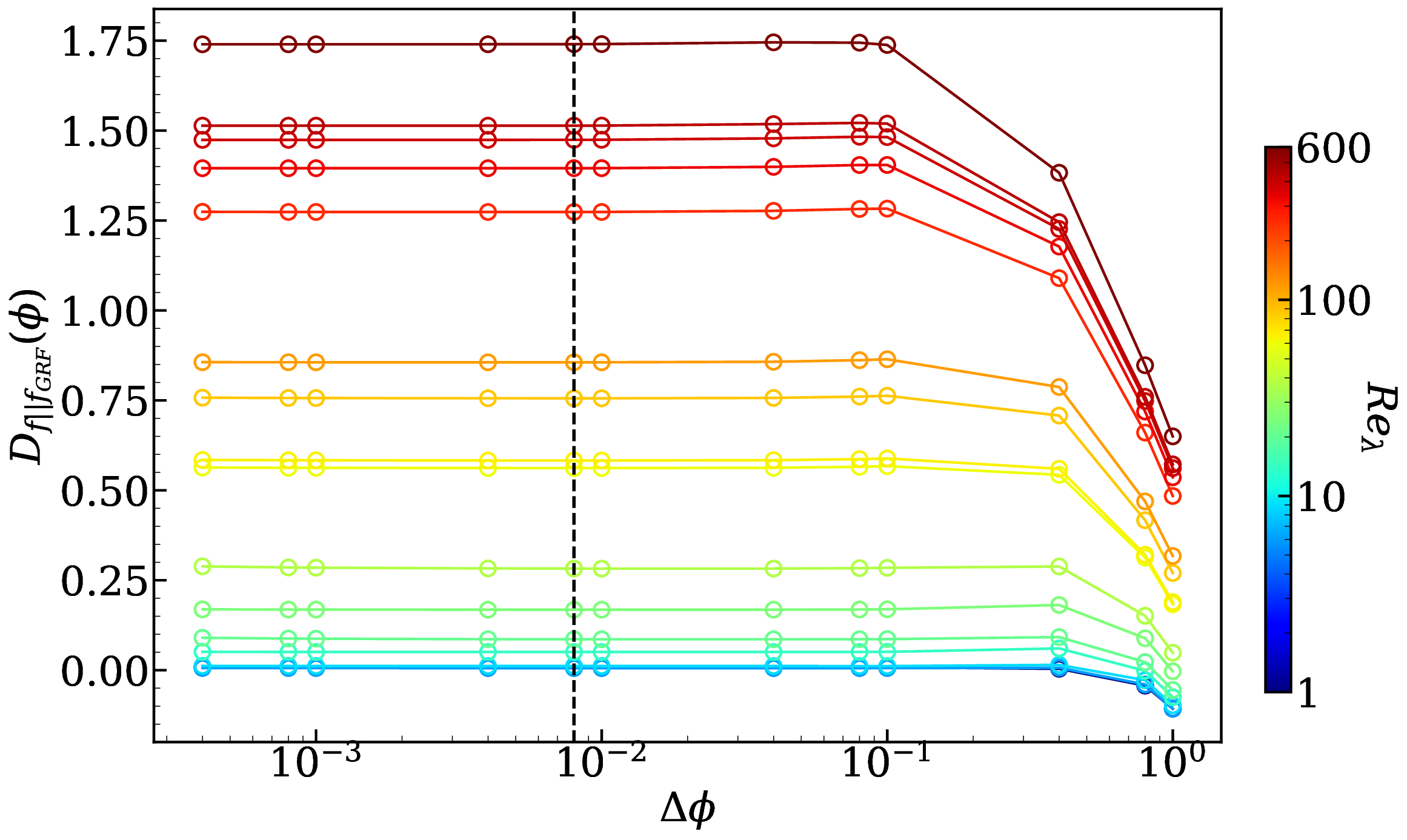}
    \caption{Variation of KL divergence of $\phi$ with bin width $\Delta \phi$ for different Reynolds numbers. Dashed line indicates the chosen bin width $\Delta \phi = 0.008$ used for all the results of this study.}
    \label{fig:convergence_KLD}
\end{figure}

To demonstrate the sensitivity of KL divergence, we perform a numerical experiment using the analytical example of a stretched exponential PDF (introduced in section \ref{sec:Analytical_interm}) representative of turbulence small-scale statistics.
We generate two sets of samples of various sizes -- one follows a stretched exponential PDF,
\begin{eqnarray}
    f_s(x) = Zx^6\exp (-bx^c)
    \label{eq:stretched2}
\end{eqnarray}
with $c=0.5$ and $b$ prescribed such that $x$ has unit mean, and the other follows a gamma distribution $f_{GRF}(x)$ with $n=8$ (Eq.~\ref{eqn4}).
The KL divergence is estimated from these data samples using the histogram approach, and plotted in Fig.~\ref{fig:KLD_sensitivity_N_delx_gamma} as a function of bin widths, for different sample sizes.
For comparison, the figure also shows the corresponding analytically obtained value of KL divergence given by,
\begin{eqnarray}
    D_{f_s||f_{GRF}}(x) = \log\left(\frac{c\Gamma\left(\frac{8}{c}\right)^7}{\Gamma\left(\frac{7}{c}\right)^{8}}\frac{\Gamma\left(\frac{n}{2}\right)}{\left(\frac{n}{2}\right)^{\frac{n}{2}}}\right) - \left(\frac{7}{c}-\frac{n}{2}\right) + \frac{7-\frac{n}{2}}{c}\left(\psi\left(\frac{7}{c}\right) - c\log\left(\frac{\Gamma\left(\frac{8}{c}\right)}{\Gamma\left(\frac{7}{c}\right)}\right)\right),
\end{eqnarray}
where $\psi(x)$ is the digamma function.
The KL divergence estimate converges to its analytical value over a range of bin widths, 
and this range widens as the sample size ($N$) increases.
The estimated value deviates from the theoretical value at very large bin widths due to an insufficient number of bins to adequately resolve the probability density function. 
Conversely, excessively small bin widths also lead to a deviation of the estimated value, due to inadequate number of samples per bin resulting in incorrect estimation of the probability density at the center of each bin. Consequently, the KL divergence estimate at lower bin widths improves with increasing sample size, expanding the range of bin widths allowed for a reliable estimate. This example demonstrates the necessity of optimal bin resolution for estimating the KL divergence of a heavy-tailed PDF accurately from a given sample size. 

\begin{figure}
\begin{tikzpicture}
\node[anchor=south west, inner sep=0] (img) at (0,0)
{\includegraphics[width=0.8\linewidth]{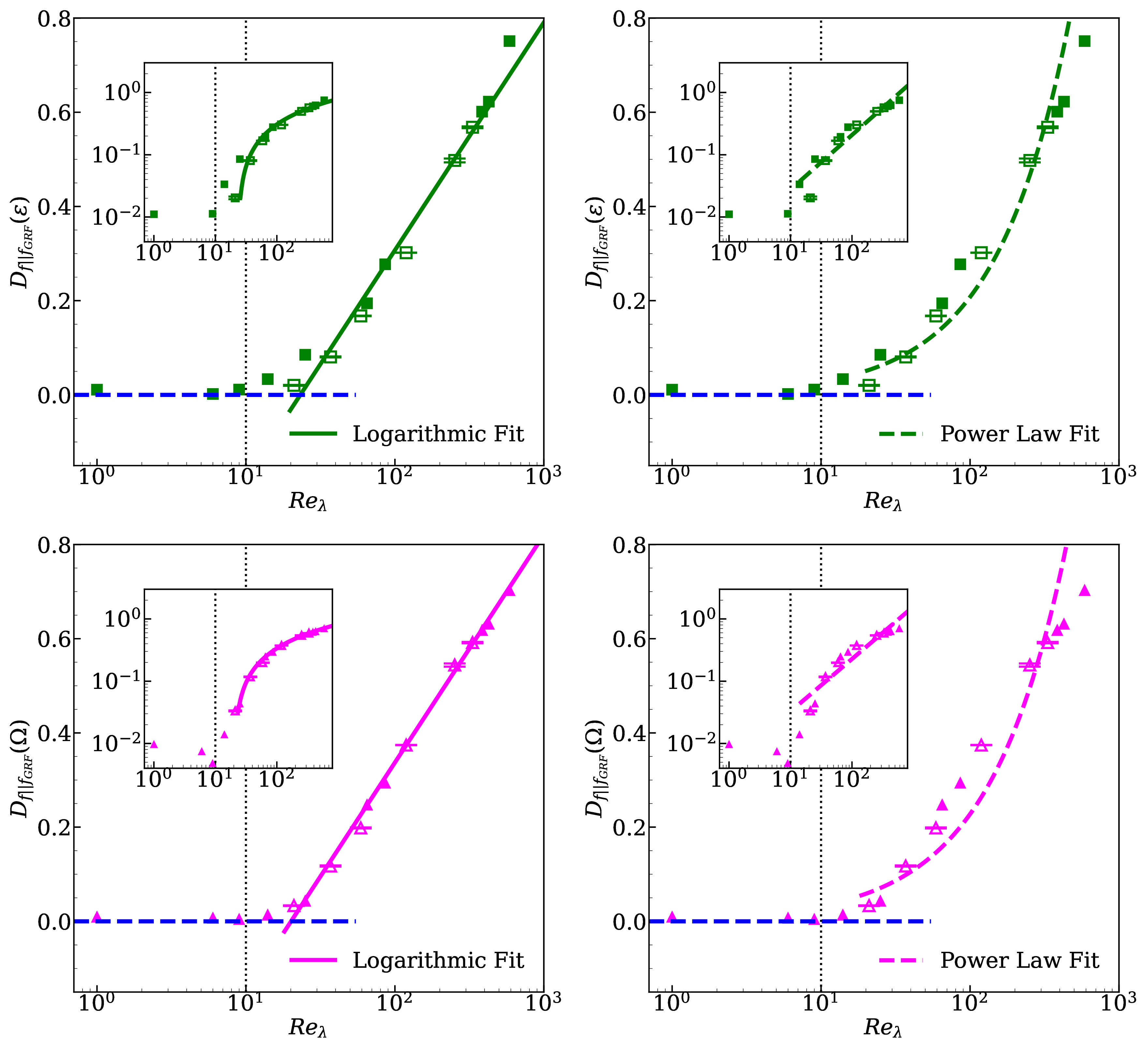}};
\begin{scope}[x={(img.south east)}, y={(img.north west)}]
          \node at (0.0, 1.0) {\textbf{(a)}};
          \node at (0.51, 1.0) {\textbf{(b)}};
          \node at (0.0, 0.49) {\textbf{(c)}};
          \node at (0.51, 0.49) {\textbf{(d)}};
    \end{scope}
\end{tikzpicture}
    \caption{KL divergence of dissipation rate $D_{f||f_{GRF}}(\epsilon)$ (green squares) and enstrophy $D_{f||f_{GRF}}(\Omega)$ (magenta triangles) as a function of $Re_{\lambda}$ in log-linear scale (log-log scale in inset). Blue dashed line indicates constant fit  $D_{f||f_{GRF}}(X) \approx 0$ for $Re_{\lambda} \leq 10$. Following fits for $Re_{\lambda} \in (20, 600)$ are reported: (a) log-law fit  $D_{f||f_{GRF}}(\epsilon)\approx -0.66 + 0.21\log{\left(Re_{\lambda}\right)}$ $(R^2 = 0.980)$, (b) power law fit $D_{f||f_{GRF}}(\epsilon)\approx 0.0037Re_{\lambda}^{0.875}$ $(R^2 = 0.884)$, (c) log-law fit $D_{f||f_{GRF}}(\Omega)\approx -0.63 + 0.21\log{\left(Re_{\lambda}\right)}$ $(R^2 = 0.997)$, and (d) power law fit $D_{f||f_{GRF}}(\Omega)\approx 0.0047Re_{\lambda}^{0.842}$ $(R^2 = 0.883)$ .}
    \label{fig:eps_ens_log_power}
\end{figure}

Next, we examine the sensitivity of the KL divergence of turbulent fields to ensure the validity of our estimates from DNS data across Reynolds numbers.
Fig.~\ref{fig:convergence_KLD} illustrates the variation of the estimated KL divergence of pseudodissipation rate $(D_{f||f_{GRF}}(\phi))$ with bin width $\Delta \phi$, over the entire range of $Re_{\lambda}$.
The estimation of KL divergence based measure of intermittency is significantly robust over a wide range of bin sizes spanning nearly three decades.
It shows deviation only at very large bin sizes, due to insufficient resolution to adequately capture the shape of the PDF.
Note that the deviation at lower bin sizes is not observed for the range considered but may be observed if the bin width is reduced further. 
The optimal bin width $\Delta \phi = 0.008$ has been chosen well within the constant range, such that KL divergence remains nearly unchanged with a small change in the bin width for all $Re_{\lambda}$.

\section{Logarithmic vs. power law scaling of dissipation and enstrophy} \label{appC}

As discussed in section \ref{sec:interm_results}, both dissipation and enstrophy show monotonically increasing turbulence intermittency with Reynolds number in the range $Re_{\lambda} \in (20,600)$. 
Fig.~\ref{fig:eps_ens_log_power} shows the qualitative comparison between 
the logarithmic (a,c) and power-law (b,d) fits over this range of $Re_{\lambda}$. It is evident that the variation of KL divergence of both dissipation and enstrophy with Reynolds number follows a logarithmic law and not a power law scaling. 
Even quantitatively, the $R^2$ scores of these fits (reported in figure caption) clearly indicate that the growth of turbulence intermittency is more accurately captured by a logarithmic scaling than by a power-law scaling.

\bibliography{apssamp}

\end{document}